\documentclass[]{spie}  

\usepackage{amsmath,amsfonts,amssymb}
\usepackage{graphicx}
\usepackage{caption}
\usepackage{placeins}
\usepackage{subcaption}
\usepackage[colorlinks=true, allcolors=blue]{hyperref}

\title{End-to-end design framework for compressed on-chip pixel-wise spectro-polarimeters}

\author[a]{T.A. Stockmans}
\author[a]{F. Snik}
\author[b]{J.M. Smit}
\author[b]{J.H.H. Rietjens}
\author[c]{M. Esposito}
\author[c]{C. van Dijk}
\author[a, d]{C.U. Keller}
\affil[a]{Leiden University, Niels Bohrweg 2, 2333 CA Leiden, The Netherlands}
\affil[b]{SRON Netherlands Institute for Space Research, Niels Bohrweg 4, 2333 CA Leiden, the Netherlands}
\affil[c]{cosine Remote Sensing, Warmonderweg 14, 2171 AH  Sassenheim, The Netherlands}
\affil[d]{Lowell Observatory, 1400 W Mars Hill Rd, Flagstaff, AZ 86001, USA}

\authorinfo{Further author information: (Send correspondence to T.A. Stockmans)\\T.A.Stockmans: E-mail: stockmans@strw.leidenuniv.nl}

\begin{document} 
\maketitle

\begin{abstract}
Modern detector manufacturing allows spectral and polarimetric filters to be directly integrated on top of separate detector pixels. This enables the creation of CubeSat-sized spectro-polarimetric instruments that are not much larger than the detector and a lens. Redundancy inherent to the observed scene, offers the opportunity for sparse sampling in the form of not scanning all filters at every location. However, when there are fewer pushbroom steps than filters, data are missing in the resulting data cube. The missing, largely redundant data can be filled in with interpolation methods, often called demosaicers. The choice of filters and their precise layout influences the performance of the instrument after the demosaicing process. In these proceedings we describe a part of a design toolbox for both the filter layout and the optimum parameters for the reconstruction to a full spectro-polarimetric data cube. The design tool is based on training a (neural) network and jointly updating the values of the filters and demosaicer. We optimized a filter layout by training on spectro-polarimetric remote observations of the Earth acquired by SPEX airborne. This optimised filter layout could reconstruct a validation scene from five overlapping snapshots (pushbroom steps), which would take 109 pushbroom steps when measuring with a classical layout and no reconstruction. 
\end{abstract}


\section{INTRODUCTION}
\label{sec:intro}  

The first instrument that performed multiband polarimetry for remote sensing from space was the POLDER instrument. \cite{deschamps_polder_1994} Since then there have been many instruments, expanding for instance on spectral capabilities; see e.g.\ the overview by ref. \citenum{dubovik_polarimetric_2019}. For space applications, volume and mass are often the most important cost drivers and hence, when similar capabilities can be achieved at reduced mass and volume, more options for application in space become available. A first step in this direction has been made with the development of the HARP cubesat \cite{martins_harp_2018} and the SPEXone \cite{hasekamp_aerosol_2019} and HARP2 polarimeters as part of the PACE mission. These instruments are significantly smaller than many of their predecessors. Still, further miniaturization while maintaining instrument capabilities would greatly simplify their use in e.g. satellite constellations, distributed systems, and on cubestats.

An effective way to design smaller spectro-polarimetric instruments is by micro-patterning both the polarimetric and spectral filters directly on each individual detector pixel. The size reduction enabled by this production technique opens up the possibility to go to CubeSat-sized instruments with all their cost benefits. Details on the micro-patterning can be found in e.g. \citenum{maruyama_32-mp_2018}. Some examples in which this technology is put to use can be found in \citenum{tu_division_2020, mu_snapshot_2017, ma_pixelated-polarization-camera-based_2019, chen_coded_2019}.

Micro-patterned filters are not new, especially concerning color filters. RGB cameras have been working for over 45 years with these kind of filters.\cite{bayer_color_1976} To enable instruments with pixelated filters to create full images within one snapshot, algorithms have been created to reconstruct the missing information, which has been dubbed "demosaicing". These algorithms infer the unmeasured values at all pixels by making use of surrounding pixels and measured bands. These techniques have expanded beyond RGB imagers to hyperspectral imaging and spectro-polarimetry. Some examples for demosaicing for polarimetric division of focal plane images are \citenum{sargent_conditional_2020, pistellato_deep_2022, zhang_learning_2018, zhang_sparse_2018}. 

The design of a filter array consists of 2 main parts. First, there are the specific parameters of the filters themselves. For spectral filters it would be the bandwidth over which it transmits; for linear polarization filters, it is the orientation angle of the linear polarizer. Second, there is the layout of these filters over all pixels; for example, the most widespread filter array layout for RGB imaging is the Bayer pattern \citenum{bayer_color_1976}. 

We have developed a network that jointly optimizes the design of the color filters for hyperspectral imaging with a optimal form of demosaicer; see \textit{Stockmans et al.\ in prep}. In these proceedings, we show the preliminary results of the addition of a micropolarizer array to this framework for the design of a spectro-polarimetric imager. 

\section{Methods}
The work presented here explores the addition of linear polarization measurements to our compressed sensing approach for hyperspectral imaging described in \textit{Stockmans et al.\ in prep}. The main focus is the design of a filter layout that can be directly attached on all individual pixels of a detector. This instrument can take a snapshot image and move the instrument slightly so that every pixel is looking at the next patch of land or ocean. We will refer to these movements as pushbroom steps and put them in as a design parameter. 

In the paper by \textit{Stockmans et al.\ in prep} we describe the layout design process of the spectral filter array on top of a CMOS detector of a hyperspectral instrument, with a demosaicing algorithm (linear reconstructor) being optimized simultaneously. The spectral filter array is designed to be incomplete, which means that every geometrical pixel is not imaged in every desired spectral band. The linear reconstructor acts as an interpolator between spectral bands and spatial connections to give an intensity estimate of every geometric pixel in all desired spectral bands. 

The design framework is built up in TensorFlow, a Python module capable of doing automatic backward propagation. The spectral filter array, detector, and reconstructor are all described as separate layers through which the data propagates from input datacube to output datacube. A layer is a TensorFlow class comprising of an input, output and updating function describing the relation between in- and output. The spectral filter array has as an input the full hyperspectral scene and the variables of the layer describe the filters themselves. The output of this layer is the filtered hyperspectral datacube. The reconstructor is a matrix of variable multipliers mapping the intensities coming out of a detector towards a full hyperspectral datacube. The variables of both layers are updated during training stage towards the values that result in the smallest difference between the reconstructed and original datacube. This difference is the so-called "loss function", a function of merit that needs to be minimized by the optimizing algorithm inherent to TensorFlow. The updating of the variables is done using back propagation, so a partial differential must be defined between the final loss function and the variables. Each variable could also be fixed to a certain value if desired, without losing the updating capabilities of the rest of the variables.

The main addition described in this proceeding to the framework above, is the addition of one layer at the beginning that describes a microgrid structure of linear polarizers in different orientations. The input of the network can be either the combination of the degree of linear polarisation (DoLP), angle of linear polarisation (AoLP) and radiance (rad) for all wavelengths, or the combination of the Stokes parameters I, Q and U for each wavelength. The intensity that falls on the simulated detector is calculated with the following equations on a pixel to pixel basis:

\begin{equation}
    I_{det} = rad\left[\frac{1-DoLP}{2}+DoLP\times \cos^2(AoLP-\theta)\right]
\end{equation}
in the case of DoLP, AoLP and rad. Or 
\begin{equation}
    I_{det} =\frac{I + Q \cos(2\theta)+U \sin{2\theta}}{2}
\end{equation}

in the case of Stokes parameters. In both equations, $\theta$ is the orientation angle of the polarizer at that pixel and $I_{det}$ is the light intensity that propagates along the network.

After the polarizer layer the datacube is manipulated further by the spectral filters and summed at the detectors to one intensity measurement per pixel. The detector layer also adds gaussian noise to mitigate overfitting and simulate a real detector closer to reality. This sequence of polarisation filters, spectral filters and detector can be repeated for every pushbroom step, before all data is multiplied by the linear reconstructor to retrieve the full reconstructed datacube.

\section{Preliminary Results}
To see the performance of our design, we have made use of the spectro-polarimetric data gathered by the SPEX airborne instrument during the ACEPOL campaign \cite{knobelspiesse_aerosol_2020}. For illustrative purposes, we trained the network to design a 5 step pushbroom imager. The spectral filter layout could be updated by the network, but the polarisation filters were kept constant in a regular 2x2 layout with alternating 0, 45, 90, 135 degree orientation of the linear polarizers. 

\begin{figure}[h]
    \begin{subfigure}[h]{0.49\textwidth}
    
        \centering
        \includegraphics[width = \textwidth]{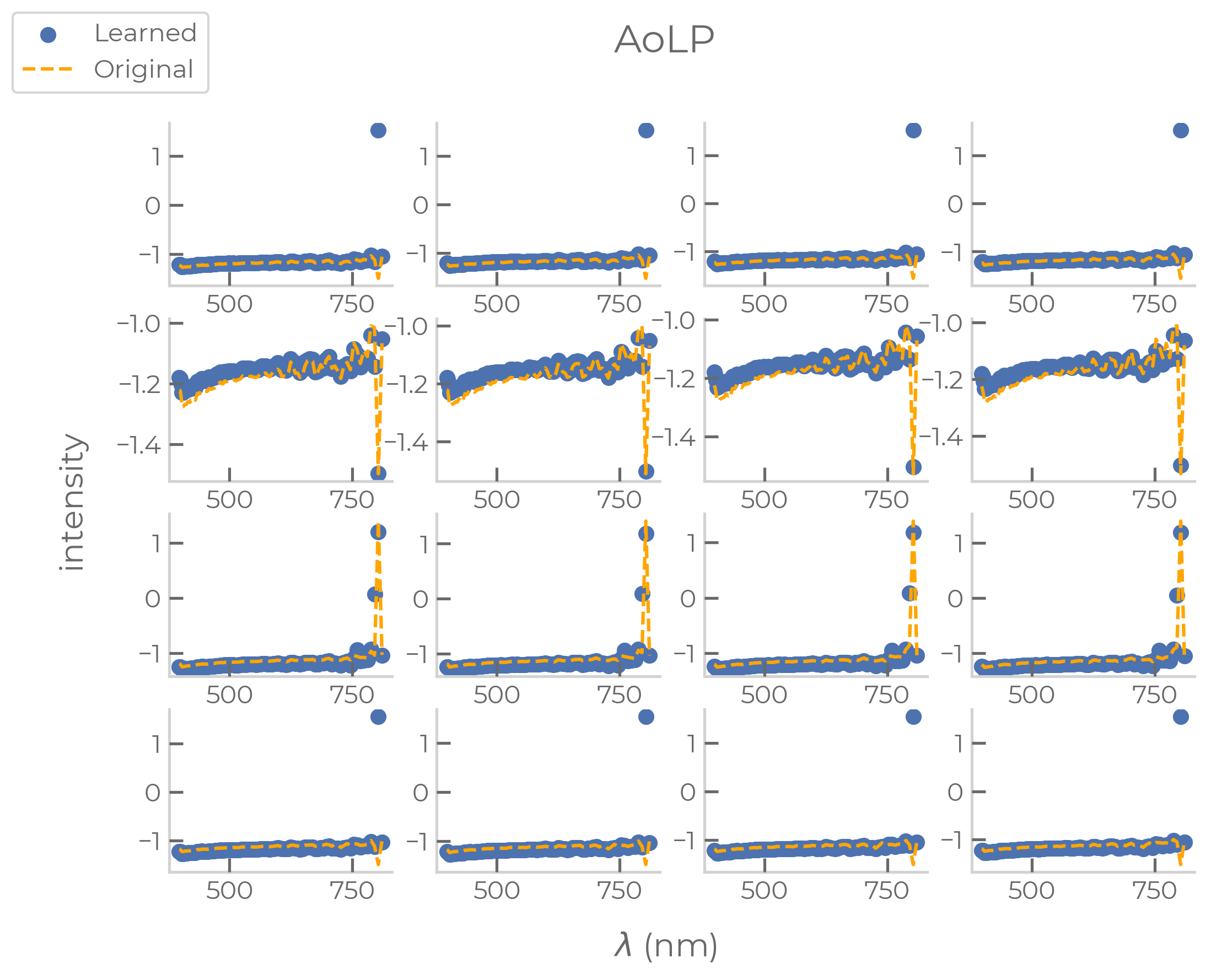}
        \caption{Angle of Linear Polarisation}
        \label{fig:reconstructed_AoLP}
    \end{subfigure}
    \begin{subfigure}[h]{0.49\textwidth}
        \centering
        \includegraphics[width = \textwidth]{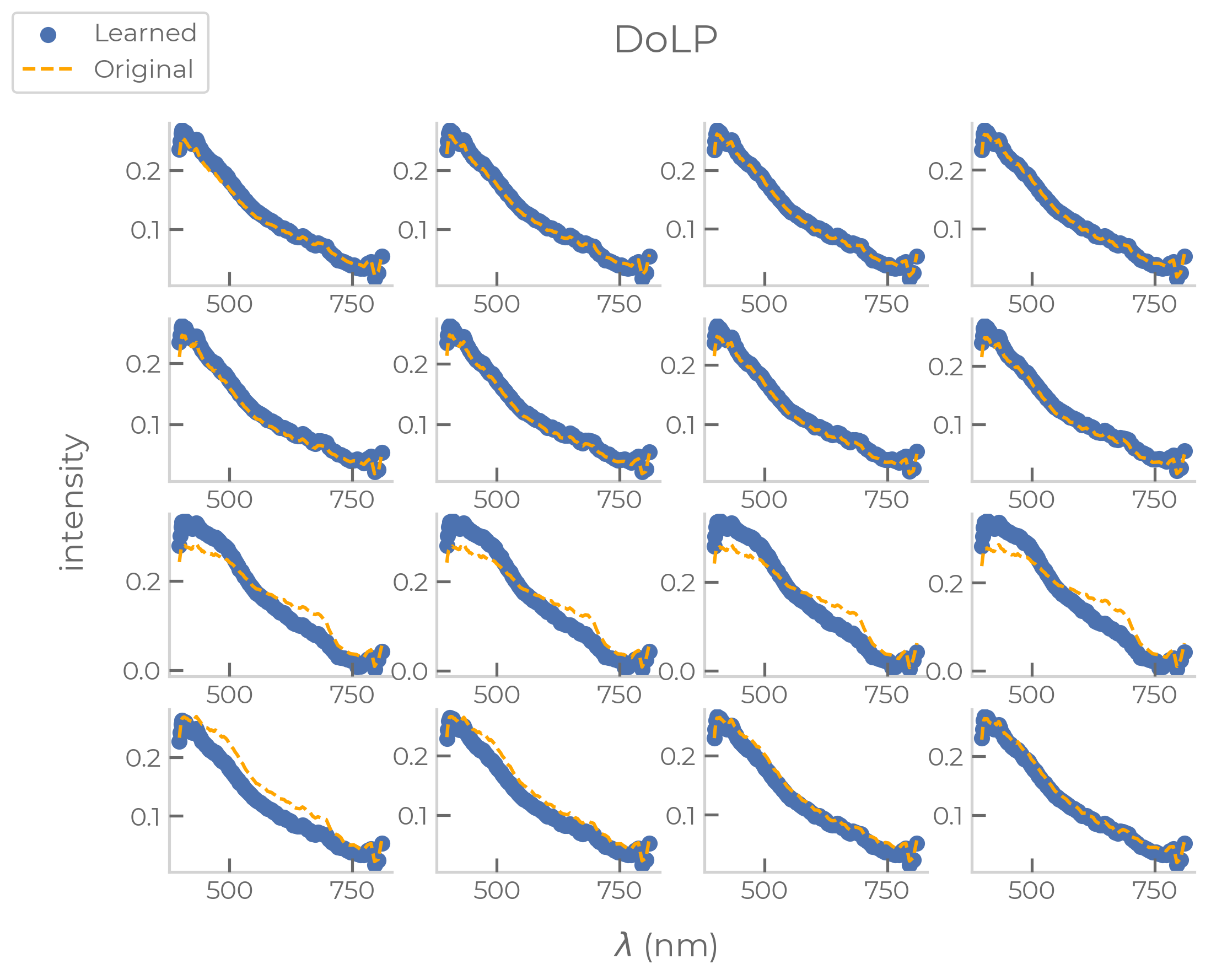}
        \caption{Degree of Linear Polarization}
        \label{fig:reconstructed_DoLP}
    \end{subfigure}
    \caption{Comparison between the original (orange dashed line) and reconstructed (blue dots) values for 16 random geometrical pixels from the SPEX airborne datacube not seen during training.}
\end{figure}

\begin{figure}[b]
    \centering
    \includegraphics[height = 16 cm]{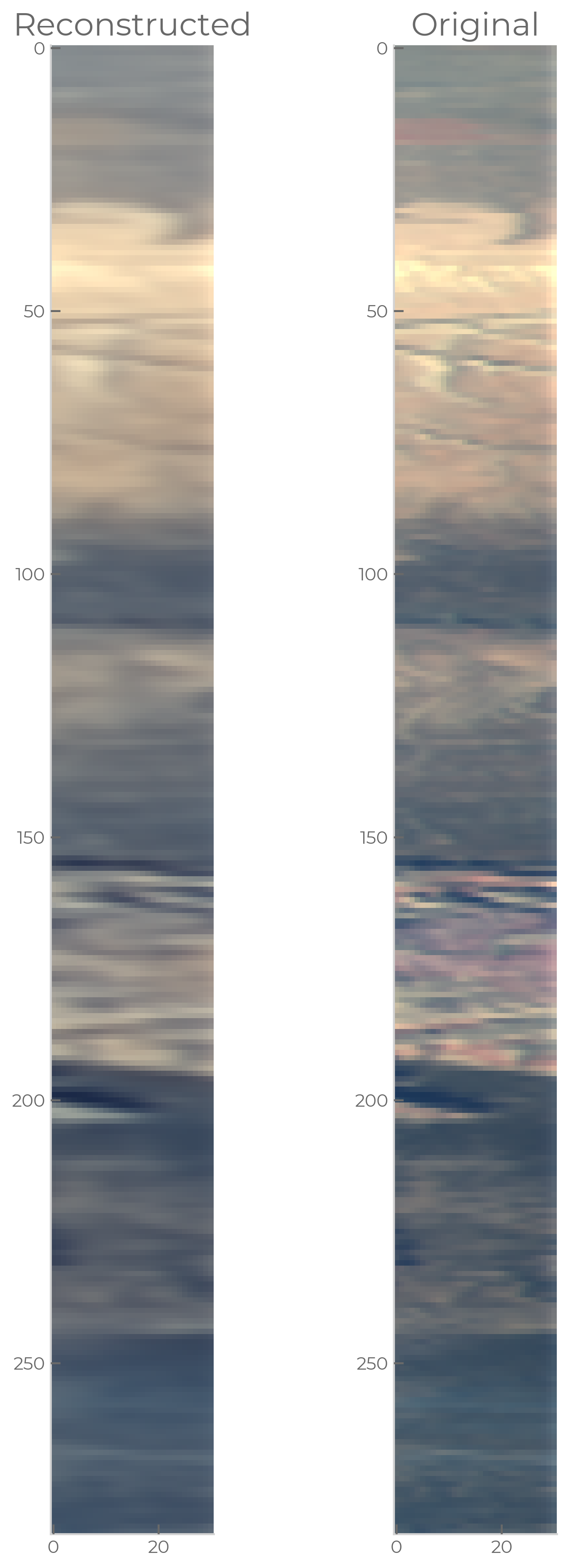}
    \caption{RGB representation of the reconstructed radiance images from a SPEX airborne datacube compared to the original.}
    \label{fig:reconstructed_images}
\end{figure}

More linescans/pushbroom steps would introduce a heavier datalink, longer acquisition time and higher power usage in order to get a more accurate image, while less linescans would have the opposite effect. The fixed layout of the polarizers was chosen to simulate the available commercial micropatterned polarimeters. The training was conducted on 10000 image patches of 10 by 10 pixels of the varying scenes (mountains, flatland, ocean, etc.) that were covered in the ACEPOL campaign. This resulted in a 10x10 spectral filter design and a corresponding linear reconstructor for the demosaicing to the full 6 dimensional datacube. We validated the performance on a scene outside of the training set. In figures \ref{fig:reconstructed_AoLP} to \ref{fig:reconstructed_images} the results are shown. It is clear that the general trends are followed well. However, some details and colors are off compared to the original. We have been able to reconstruct the radiances with a SNR of 24.9 dB. The AoLP are on average 2.6 degrees off, and the DoLP carries a mean error of 0.018.

\section{Discussion and conclusion}
The pixelated filter instruments that can be designed with our toolbox are expected to find their best use in a monitoring function. It would be harder to collect and fully characterize an entirely new spectrum with these instruments, since the demosaicing is trained on known spectra. However, when used next to the existing instruments, these instruments would be the cheaper option to get a temporal monitoring of some changing processes.

The two main variable parameters that can be changed for each design is the number of pushbroom steps that should be taken and the amount of filters that the network should converge to. The choice of these two parameters effects the size of the datalink, retrieved accuracy, operation costs, and production feasibility. 
The described toolbox has a lot of flexibility. In this proceeding we made use of the simplest possible reconstructor for the demosaicing. However, the combination with different and more intricate demosaicing algorithms is possible as long as it is compatible with the TensorFlow API. Also, for more specified end products, the loss function could be adapted to make a specified design for a certain function of merit or post-processing algorithm.

\FloatBarrier
\bibliography{SPIE2022} 
\bibliographystyle{spiebib} 

\end{document}